\documentclass{article}
\usepackage{ltwol2e}

\usepackage{epsfig}
\usepackage{color,latexsym} 

\def\Journal#1#2#3#4{{#1} {\bf #2}, #3 (#4)}

\def\NPB{{\em Nucl. Phys.} B}
\def\PLB{{\em Phys. Lett.}  B}
\def\PRL{\em Phys. Rev. Lett.}
\def\PRD{{\em Phys. Rev.} D}

\def\be{\begin{equation}}
\def\ee{\end{equation}}
\def\bea{\begin{eqnarray}}
\def\eea{\end{eqnarray}}

\begin{document}

\title{
\vspace*{-1.5cm}
\begin{flushright}
{\small KANAZAWA 98-16, HUB-EP 98/63}
\end{flushright}
\vspace*{0.5cm}
GAUGE INVARIANT PROPERTIES OF ABELIAN MONOPOLES}

\author{S. THURNER, H. MARKUM}

\address{Institut f\"ur Kernphysik, TU-Wien, 
Wiedner Hauptstra\ss e 8-10, A-1040 Vienna  
\\E-mail: thurner@kph.tuwien.ac.at}   

\author{E.-M. ILGENFRITZ} 
\address{Institute for Theoretical Physics, Kanazawa University, Japan}
\author{ M. M\"ULLER-PREUSSKER}
\address{Institut f\"ur Physik, Humboldt Universit\"at zu Berlin, Germany  }

\twocolumn[\maketitle\abstracts{
Using a renormalization group motivated smoothing technique, 
we investigate the large scale structure of lattice
configurations at finite temperature, concentrating on 
Abelian monopoles identified in the maximally Abelian, the Laplacian Abelian, 
and the Polyakov gauge. 
Monopoles are mostly found in regions of large 
action and topological charge, rather independent of the 
gauge chosen to detect them. 
Gauge invariant properties around Abelian monopoles, 
the local non-Abelian action
and topological density, are studied. 
We show that the local averages of these densities 
along the monopole trajectories are clearly above the background, which 
supports the existence of monopoles as physical objects.
Characteristic changes of the vacuum structure at the deconfinement
transition can be attributed to the corresponding Abelian monopoles,
to an extent that depends on the gauge chosen for Abelian projection.
All three Abelian projections reproduce the full $SU(2)$ string tension
within $10$ \% which is preserved by smoothing.
}]

\section{Introduction}

\vspace{-2mm}
Over the last two decades a variety of attempts 
in field theory have been aiming for a qualitative understanding and modeling 
of two basic properties of QCD: quark confinement and 
chiral symmetry breaking. The most prominent schemes are the instanton 
liquid model \cite{ILM} and the dual superconductor picture of the 
QCD vacuum.\cite{DSC}
While the first model explains chiral symmetry breaking and solves the $U_A(1)$ 
problem, the second 
one provides a simple idea for the confinement mechanism. 
In this scenario, where the vacuum is viewed as a dual superconductor,   
condensation of color magnetic monopoles leads to confinement 
of color charges through a dual Meissner effect. 
The superconductor picture was substantiated by a large number of 
lattice simulations over the last years. So it was shown that 
in the confinement phase monopoles percolate through the $4D$ 
volume \cite{BORN} 
and are responsible for the dominant contribution to the string 
tension.\cite{SUZU} 
At present, more and more groups  
characterize their lattice vacuum in accordance to the instanton liquid
picture.\cite{NEGELE98}

Both models rest on the existence of very different 
kinds of topological excitations,
instantons and color magnetic monopoles. 
For a long time they have  been treated 
independently, only recently some deeper connection among those different 
objects has been pointed out,
both on the lattice and in the continuum.\cite{MONINST}   
Instantons are localized solutions of the  Euclidean equations of motion 
in Yang-Mills theory carrying action and integer topological charge. 
Even though it is difficult to detect instantons and antiinstantons among 
quantum fluctuations, there is no problem to study these well-defined 
objects in classical 
or semiclassical (heated) configurations on the lattice. 
The situation for monopoles is more difficult. 
Following 't Hooft, monopoles should be searched for 
as pointlike singularities of some gauge transformation dictated by a local, 
gauge 
covariant composite field. The standard prescription, however, is localizing 
monopoles in QCD as Abelian monopoles via an Abelian projection from some
gauge (for example the maximally Abelian gauge). This leads to monopole 
trajectories which are dependent on the gauge chosen.
Note however, that the condensation mechanism of monopoles itself seems to be 
gauge independent.\cite{DIGI2} \\
In this contribution we relate gauge invariant observables to monopole 
trajectories, with the intention to further understand the semiclassical 
vacuum structure in terms of monopoles and lumps of topological charge 
and their role for the confinement problem. 
We will comment on a new way of monopole identification on the lattice 
which evades serious problems of previous methods, and which might have 
a close formal relationship to 't Hooft-Polyakov monopoles. 

\vspace{-1mm}
\section{Smoothing} 

\vspace{-1mm}
To resolve semiclassical structures in gauge field configurations
provided by lattice simulations, the cooling method has been used, 
which locally minimizes the action. 
However, even improved versions of cooling rapidly destroy monopole 
percolation and reduce the string tension.
From the instanton point of view cooling is known to destroy small 
instantons and instanton-antiinstantons pairs, such that the true 
topological structure is accessible at best by a backward 
extrapolation to zero cooling steps.
Up to now, most lattice studies are performed with the Wilson action, 
for which the lattice definitions of the topological charge $Q$ are 
known to violate the bound valid for the continuum action: 
$S\geq 8\pi^2|Q|/g^2$, 
where $g$ is the coupling constant. The Wilson action is known to 
decrease with smaller size $\rho$ of an instanton, such that 
isolated instantons are 
unstable under cooling. In contrast to this, improved  
cooling finds instantons stabilized within a size interval 
$\rho > 2~a$ ($a$ is the lattice spacing). Other methods like APE smearing
let instantons grow.

\begin{figure} 
\vspace{-0.3cm}
\begin{tabular}{cc}
\hspace{-3.0mm} \epsfxsize=4.5cm\epsffile{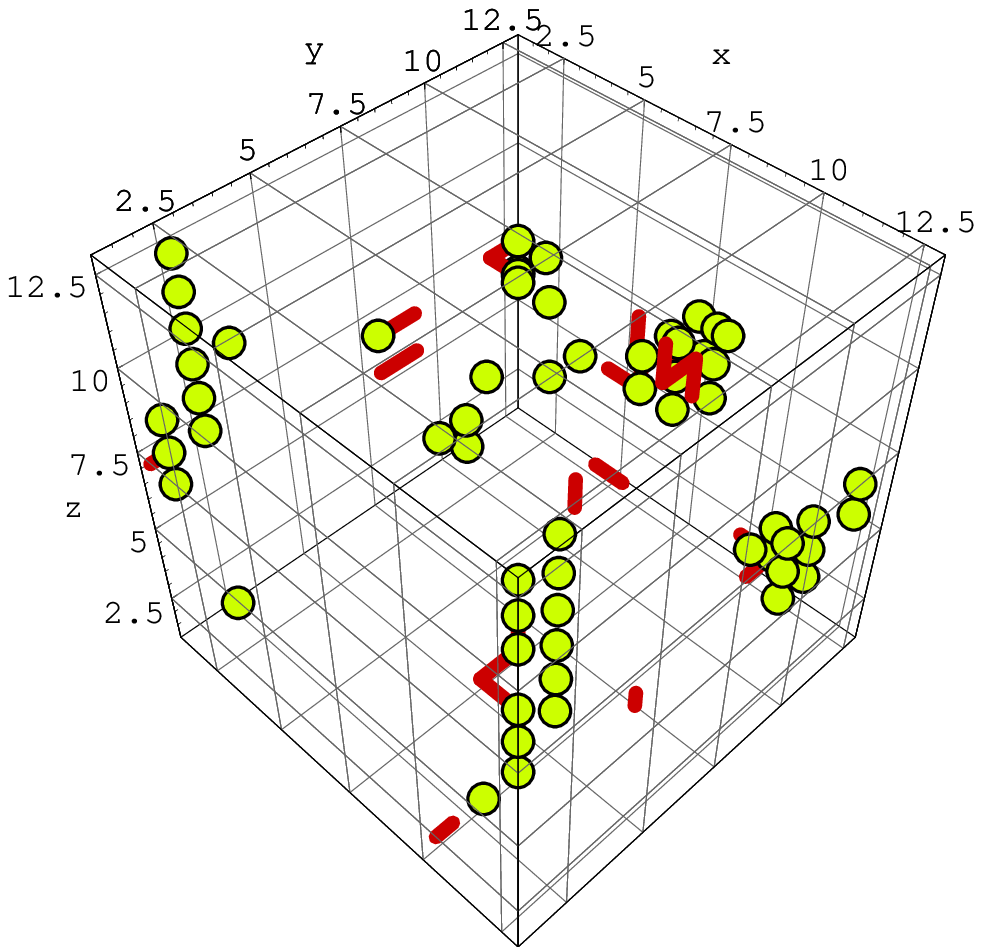}&
\hspace{-9mm} \epsfxsize=4.5cm\epsffile{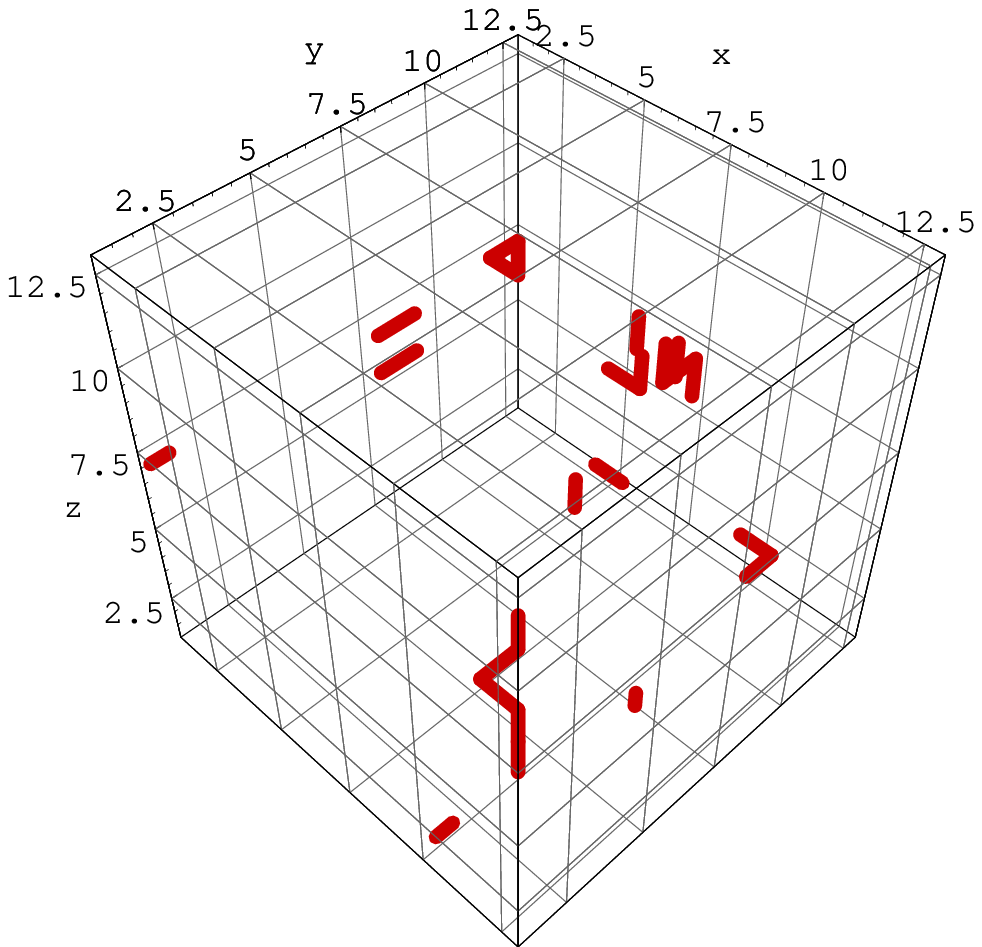}
\end{tabular}
\caption{Regions of low modulus of the auxiliary Higgs field (dots), 
	 which {\it should} mark the trajectories of monopoles according to
	 LAG, are found very close to the trajectories of DGT monopoles 
         obtained by Abelian projection. For clarity DGT monopoles are 
         also shown alone (right).
}
\vspace{-0.1cm}
\label{fig1}
\end{figure}
To avoid these ambiguities we have used a method of `constrained 
smoothing' \cite{FEUE} which is based on the concept of perfect 
actions.\cite{HASE} 
These actions respect the above bound for the topological charge
and lead to a theoretically consistent `inverse blocking' 
operation. 
Inverse blocking is a method to find a smooth interpolating field on a 
fine lattice by constrained minimization of the perfect action, provided
a configuration is given on a coarse lattice.
This makes an unambiguous definition of topological charge possible.  
Constrained smoothing is a renormalization group motivated method  
which first blocks fields $\{U\}$, sampled on a fine lattice with lattice 
spacing $a$, to a coarse lattice $\{V\}$ with lattice spacing $2~a$ 
by a standard blockspin transformation. Then inverse blocking is used to 
find a smoothed field $\{U^{\mathrm{sm}}\}$ replacing $\{U\}$.

An important feature of this method is that it 
does not drive configurations into classical fields as unconstrained
minimization of the action would do. It saves the long range structure of the 
Monte Carlo configuration in $\{V\}$, such that the smooth background 
contains semiclassical objects deformed by classical and quantum interaction. 
The upper blocking scale roughly defines the border line between 'long and 
short range'\footnote{The iterative application of this method, 
`cycling',\cite{BOUL} obscures the idea of a definite blocking scale while 
it still preserves rather well features of long range physics as the string 
tension.}.
In this work we used a simplified fixed-point action \cite{PRD98} for Monte
Carlo sampling and for constrained smoothing before the configurations 
were analyzed.
\begin{figure}[thb] 
\epsfxsize=8.5cm\epsffile{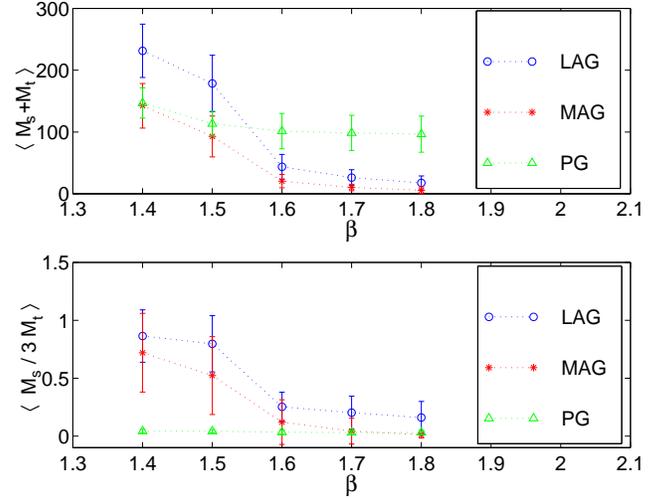}
\vspace{-2mm}
\caption{Total monopole length (top) and space-time asymmetry (bottom) 
         as a function of $\beta$ for monopoles obtained in different gauges. 
         $\beta_c = 1.545(10)$ is the deconfinement point.  }
\label{fig2}
\vspace{-1mm}
\end{figure}

\section{Gauge Fixing}

The most popular gauge to study monopoles on the lattice 
is the maximally Abelian gauge (MAG).\cite{KRON87} This gauge is enforced by 
an iterative minimization procedure, which can get stuck in local minima,  
so-called {\it technical} Gribov copies. The Laplacian Abelian gauge 
(LAG) \cite{SIJS} is not afflicted by this problem.
MAG and LAG can be understood along the same lines.  
The gauge functional of the MAG can be written: 
\begin{eqnarray}
F(\Omega) & = &\sum_{x,\mu} (1-\frac{1}{2} \, tr \, 
          (\sigma_3 U^{(\Omega)}_{x,\mu} 
           \sigma_3 U^{(\Omega)\dag}_{x,\mu} ) ) \nonumber \\
= & &  \sum_{x,\mu,a} (X_x^a - \sum_{b} R^{a,b}_{x,\mu}
          X_{x+\hat \mu}^b )^2 \rightarrow \int_V (D_{\mu} X)^2 \, ,
\end{eqnarray}
with the gauge transformation $\Omega_x$ acting on $\{U\}$
\begin{displaymath}
U^{(\Omega)}_{x,\mu}  = \Omega_x U_{x,\mu} \Omega^{\dag}_{x+\hat \mu}
\end{displaymath}
encoded in an {\it auxiliary} adjoint Higgs field
\begin{displaymath}
\Phi_x  =  \Omega^{\dag}_x \sigma_3 \Omega_x = \sum_a X^a_x \sigma_a 
\end{displaymath}
subject to local constraints $\sum_a (X^a_x)^2 =  1$ 
and with adjoint links
\begin{displaymath}
R^{a,b}_{x,\mu}  =  \frac{1}{2} \, tr \, (\sigma_a U_{x,\mu} \sigma_b 
                                               U_{x,\mu}^{\dag}) \, .
\end{displaymath}
In LAG the local constraints are relaxed and 
replaced by a global normalization: 
$ \sum_{x,a} (X^a_x)^2 = V $, such that 
Eq. (1) can be further written: 
\begin{equation}
\int_V (D_{\mu} X)^2 \rightarrow \sum_{x,a} \sum_{y,b}
X^a_x\{-\Box_{x,y}^{a,b}(R) \} X^b_y \, . 
\end{equation}
Then the minimization reduces to a search for  the lowest eigenmode 
of the covariant lattice Laplacian. 
LAG is unambiguously defined, except for degenerate 
lowest eigenmodes, which correspond to {\it true} Gribov copies.  
For both MAG and LAG, the gauge transformation is finally performed 
by diagonalization of the field $\Phi_x$.  
Quite similarly we enforce  the 
Polyakov gauge (PG) by diagonalization of Polyakov loops. 
\begin{figure}[thb] 
\begin{tabular}{c}
\hspace{0.4cm} \epsfxsize=8.4cm\epsffile{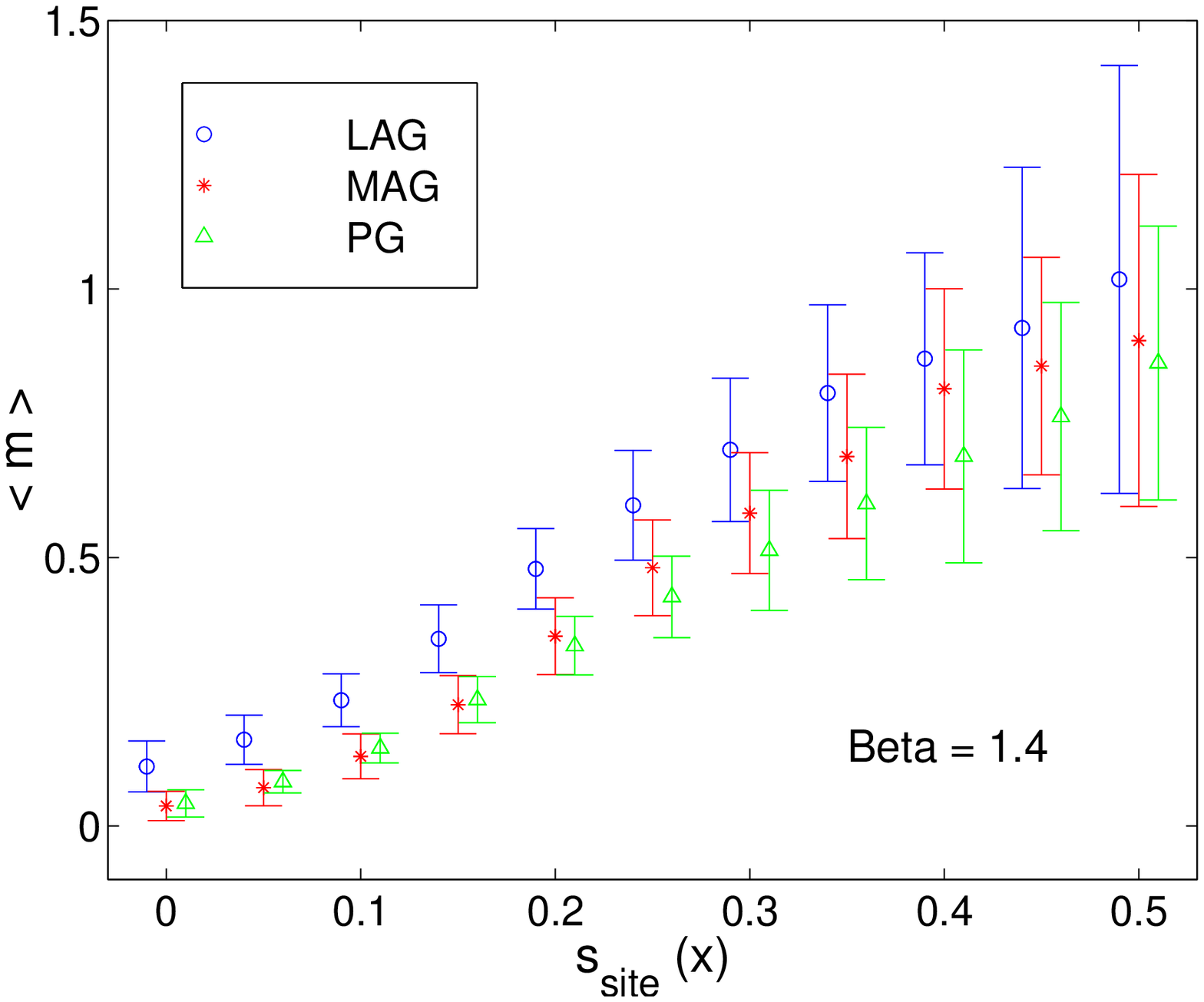}\\
\epsfxsize=8.0cm\epsffile{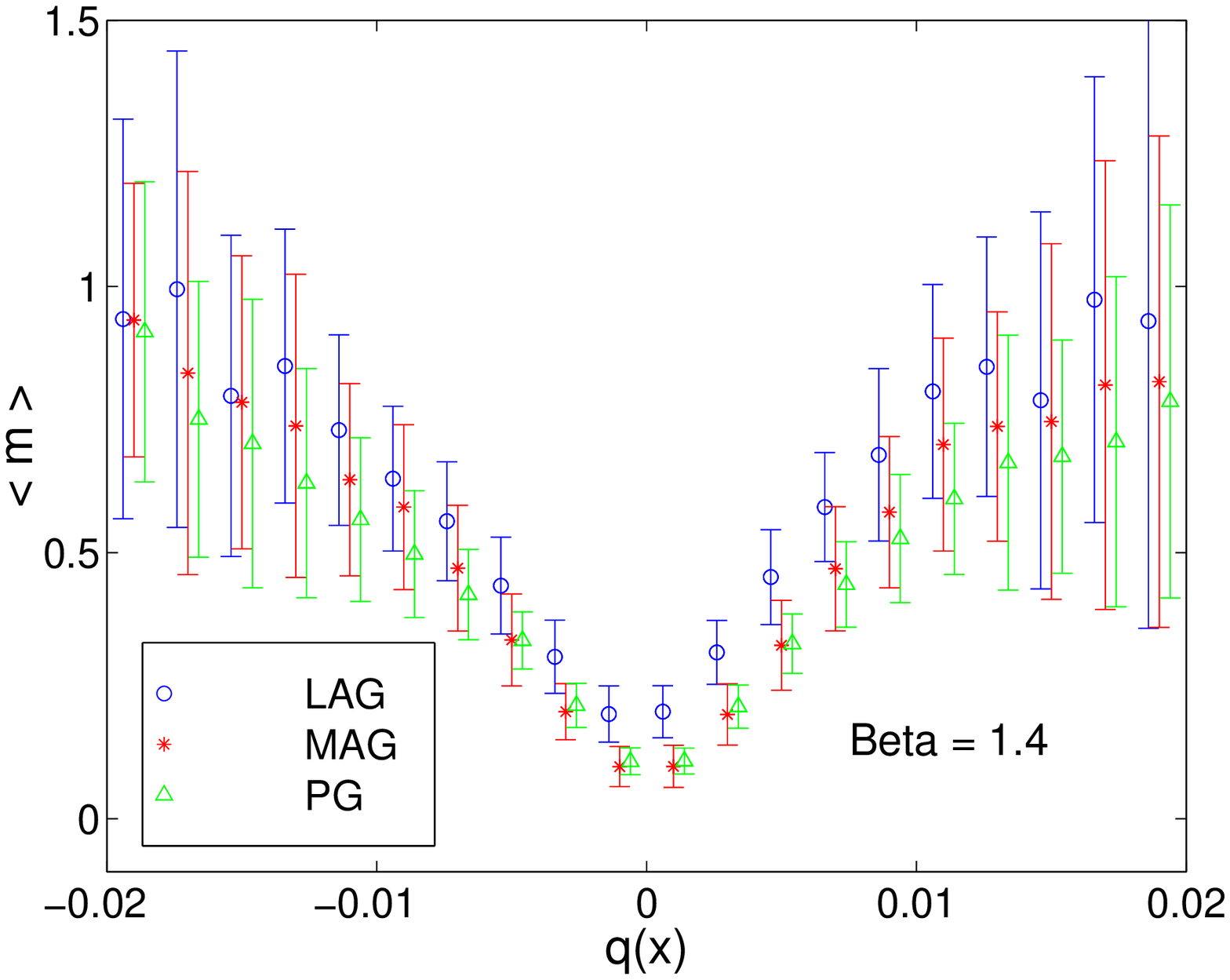}
\end{tabular}
\caption{Average occupation number of monopoles $<m>$ nearest to sites 
with action density
$s_{\rm site}$ (top) and topological charge density $q$ (bottom) in  
the confinement phase.}
\label{fig3}
\vspace{-0mm}
\end{figure}

After the Abelian gauge of choice has been fixed one extracts the Abelian 
degrees of freedom (Abelian projection). The Abelian link angles 
can then be used for the identification of monopoles, 
like in compact $U(1)$ theory, searching for the ends of Dirac strings. 
Monopoles identified in this manner are generally referred to as 
DeGrand-Toussaint (DGT) monopoles. 
The Higgs field indroduced in the LAG provides an alternative 
for monopole identification which is more satisfactory from 
a physical point of view.   
Lines of $\rho_x=|\Phi_x|=0$ where $\rho_x$ is defined as
\begin{equation}
 X^a_x = \rho_x \hat X^a_x \,\,\ , \,\,\, 
\rho_x = \sqrt{\sum_{a=1}^3 (X^a_x)^2 }  \, ,
\end{equation}
directly define lines of gauge fixing 
singularities (mo\-no\-po\-les), 
more in the original spirit of 't Hooft.\footnote{For the 't Hooft-Polyakov 
monopole regions with $\rho=0$ of the {\it physical} Higgs field are identified
with the centers of such monopoles.} Note here that this way of monopole
identification does not require to perform the actual gauge fixing and 
Abelian projection!

In Fig. 1 we show that both methods of monopole identification
turn out to be quite related. 
Regions of small $\rho$ are highly correlated with trajectories of monopoles 
identified by the DGT method. 
\begin{figure}[htb] 
\begin{tabular}{c}
\epsfxsize=7.5cm\epsffile{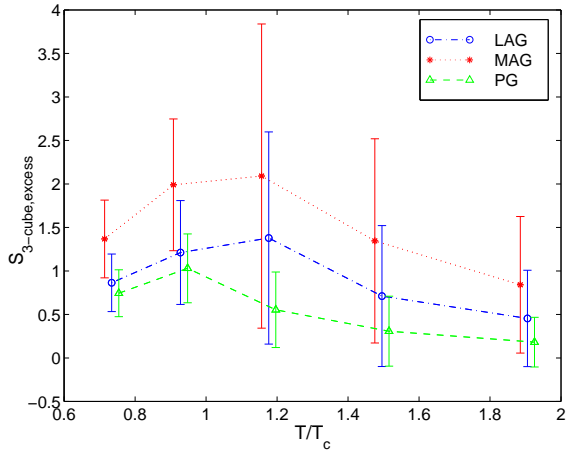}\\
\epsfxsize=7.5cm\epsffile{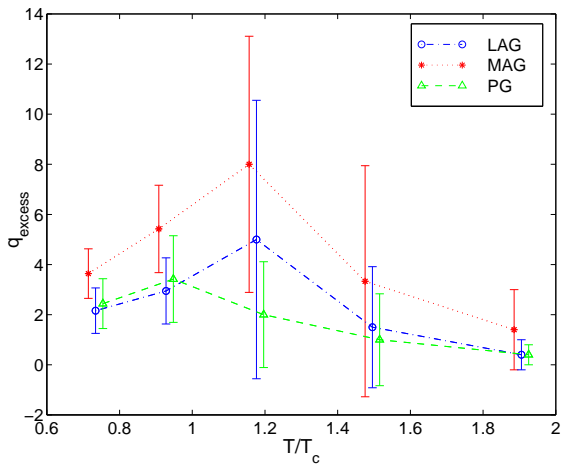}\\
\end{tabular}
\vspace{-1mm}
\caption{Excess action (top) and charge (bottom)  as a  function of 
temperature.   }
\vspace{-1mm}
\end{figure}

\section{Physical Properties of Monopoles} 

The following results were obtained from simulations  
of pure $SU(2)$ theory 
on a $12^3\times 4$ lattice. Observables were computed on 
50 independent configurations per $\beta$. 
Different $\beta$ values were considered 
to study the behavior slightly below and above the 
deconfinement phase transition. For the particular action used,\cite{PRD98}
$\beta_c=1.545(10)$. 
Global properties like 
the total loop length and the space-time asymmetry 
are shown in Fig. 2. 
DGT monopoles extracted from the MAG and the LAG behave qualitatively similar. 
Those from the the PG show no change  
at the deconfinement phase transition. This reflects the 
fact that PG monopoles should be static. 

In Fig. 3 we present the average occupation number of monopoles on dual links 
nearest to a given site as a function of the local 
action $s_{site}(x)$ and charge $q(x)$,
for different gauges.
One observes 
that the probability of finding monopoles increases 
with the amount of action/charge  density at the same lattice position.
This result is practically independent of the  gauge  used
to define the (DGT) monopoles. 

If monopoles are physical objects, one  expects that they can  
be characterized by a local excess of the (gauge invariant) action.
We define such an  excess action of monopoles by  
\begin{equation}
S_{\rm ex}= \frac{< S_{\rm monopole}-S_{\rm no monopole} > }{<S_{\rm no monopole} >}  \,\, ,
\end{equation}
where $S_{\rm monopole}$ is the action contained in a three-dimensional 
cube which corresponds to the dual link occupied by a monopole.  
Replacing the action in the above expression by the modulus of the 
topological charge density according to the L\"uscher method 
we obtain the charge excess $q_{\rm ex}$. 
For details of the definition of the local operators see Ref. 11.
Fig. 4 shows that just below $T_c$ the excess action and charge for the 
MAG and LAG monopoles are clearly above one, 
indicating an excess of action of more than a factor of two compared to the
bulk average (background). 
The large error bars above $T_c$ reflect the fact that the 
topological activity diminishes  in the deconfinement phase. 
These results are somewhat enhanced in comparison  
to  a $T=0$ study with Wilson action
without cooling or smoothing.\cite{BAKK98} 

\begin{figure}[htb] 
\vspace{-0mm}
\epsfxsize=7.5cm\epsffile{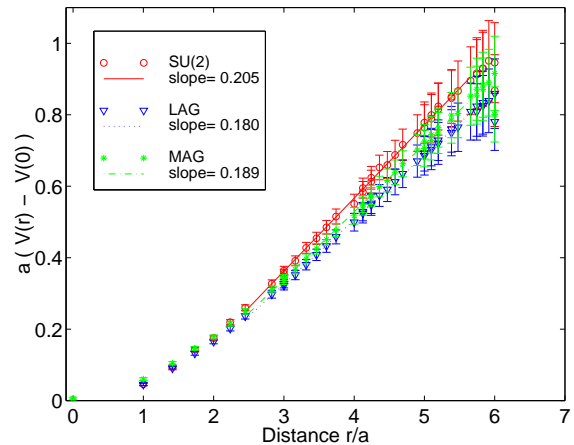}
\caption{Static quark-antiquark potentials obtained from 
Pol\-ya\-kov-line correlators after smoothing. 
The slope of the Abelian potential in the MAG 
after Abelian projection is about 5\% less than that of $SU(2)$ gauge field. 
Another 5\% are lost in the LAG. Still the LAG carries 90\% 
of the original string tension, indicating Abelian dominance 
for the LAG.}
\label{fig4}
\end{figure}

In Fig. 5 we display  static quark-antiquark potentials  obtained from 
Polyakov-Antipolyakov correlators, for the $SU(2)$ fields 
and, in the case of LAG and MAG, 
after Abelian projection. 
The Abelian string tension of the MAG is about 5\% less than that
the $SU(2)$ field. The Abelian string tension of LAG is a little smaller 
than for the MAG but still exhibits Abelian dominance. 
The Abelian string tension of PG is trivially identical with that measured
on the smoothed $SU(2)$ configurations.

Finally we present an intuitive  argument, that mono\-poles also should 
carry electric charge, that they are dyons. 
Consider the 
selfduality equations $F_{\mu \nu} = \pm \tilde F_{\mu \nu}$. 
From the trivial relation  
\begin{equation} 
\int d^4 x {\rm Tr} [(F_{\mu \nu} \pm \tilde F_{\mu \nu})^2] \geq 0
\end{equation}
it immediately follows that  
\begin{equation} 
S \geq \frac{8 \pi}{g^2} |Q|  
\end{equation}
and (anti)selfdual fields saturate the identity.
Fig. 6 depicts the probability 
distribution of topological charge density for a given local action and  
shows that for 
$s_{site}>0.3$ the local action obeys a local version of Eq. (6)
near to saturation,
$s_{site}(x) \sim \frac{8 \pi}{g^2} |q(x)|$. 
The plot was obtained after one constrained smoothing step, 
and exhibits that the gauge fields are already sufficiently smooth to 
expose semi-classical structure.  This is suggested by the relatively clear  
ridges indicating approximate local selfduality for large enough action density.
In Fig. 3 we provided evidence that  monopoles are found predominantly 
in regions of large action. 
We thus conclude that monopoles also carry electric 
charge and should be interpreted as dyons.  
Note that this way of argumentation is a shortcut, to be more precise, 
one would have to test for local selfdualtiy along individual monopole 
trajectories. 

\begin{figure}[htb] 
\vspace{-1mm}
\begin{tabular}{c}
\epsfxsize=9.0cm\epsffile{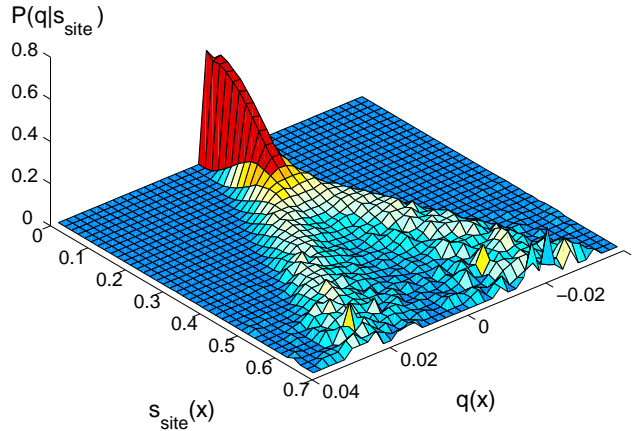}\\
\end{tabular}
\vspace{-1mm}
\caption{Probability distribution for finding a topological 
charge density $q(x)$ at a lattice site $x$ if the local action density 
equals $s_{site}(x)$. The ridges follow the lines  
$s_{site}(x) = \frac{8 \pi}{g^2} |q(x)|$ where local 
(anti)selfduality is satisfied.}
\vspace{-1mm}
\end{figure}

\section{Conclusions}

We have demonstrated that the renormalization group smoothing technique  
with an (approximate) classically perfect action provides a powerful tool 
to investigate the semiclassical vacuum structure. Analyzing trajectories 
of monopoles identified in various gauges we found that monopoles appear 
preferably in regions  which are characterized by enhanced action and 
topological charge density. We showed that in exactly those regions local 
(anti)selfduality of the gauge fields is prevailing. 
This is further evidence that monopoles should be 
addressed as dyons.  
We demonstrated that almost the complete string tension 
can be recovered from the Abelian projected field corresponding to
various Abelian gauges, indicating Abelian dominance also for the LAG. 
This is trivially true for the PG, but the corresponding
monopoles do not change at the deconfinement transition.
We have shown that monopole trajectories carry an excess action of 
about twice the background action density of smoothed gauge fields. 
Similarly, monopoles also carry excess topological charge. 
In the confinement phase this observation is rather independent of the 
gauge chosen for identifying Abelian monopoles, but the behavior of
PG monopoles is different in the deconfinement.
We therefore conclude that MAG and LAG monopoles 
behave similar physically and can be 
interpreted  as physical objects which carry action and topological charge. 
\\
\\
\noindent
This work was supported in part by FWF,  No. P11456.

\end{document}